\documentstyle[aps,epsfig,12pt]{revtex}
\title{Spin flip from dark to bright states in InP quantum dots}
\author{D.W. Snoke$^1$, J. H\"ubner$^2$, W.W. R\"uhle$^2$, and M. Zundel$^3$\\
$^1$Department of Physics and Astronomy, University of Pittsburgh\\
$^2$ Physics Department and Materials Science Center, Philipps
University, Renthof 5, 35032 Marburg, Germany\\
$^3$ Max-Planck-Institut f\"ur Festk\"orperforschung,
Heisenbergstrasse 1, 70569 Stuttgart, Germany }

\begin{document}

\maketitle
\vspace{1cm}

\begin{abstract}

We report measurements of the time for spin flip from dark (non-light emitting) exciton states in
quantum dots to bright (light emitting) exciton states in InP quantum dots. Dark excitons are
created by two-photon excitation by an ultrafast laser. The time for spin flip between dark and
bright states is found to be approximately 200 ps, independent of density and temperature
below 70 K. This is much shorter than observed in other quantum dot systems. The rate
of decay of the luminescence intensity, approximately 300 ps, is not simply equal to the radiative
decay rate from the bright states, because the rate of decay is limited by the rate of conversion
from dark excitons into bright excitons. The dependence of the luminescence decay time
on the spin flip time is a general effect that applies to many experiments.
 \end{abstract}

\newpage
There has been increasing interest in the spin flip properties of quantum dots,
especially relating to the study of ``spintronics.'' Quantum dots have also
been proposed as elements of quantum computers. One of the appeals of this system for quantum
computing applications is the observation of very long spin flip time for carriers in quantum
dots. Electron spin flip times are found to be of the order of microseconds in InGaAs dots
\cite{fuji}, while exciton spin flip times, which include hole spin flip, have been found to be at
least several nanoseconds in InAs dots \cite{paillard}.  Measurements of the dephasing rate,
which are sensitive to the total rate of all scattering processes, including spin flip, have found
dephasing times for excitons in quantum dots of hundreds of picoseconds
\cite{birkedalPRL01,borriPRL01}. 

It is tempting to view the long spin flip times observed in these sytems as an intrinsic property
of quantum dots. In quantum wells and bulk semiconductors, spin flip can occur in conjunction
with scattering between different {\bf k}-states, because the valence band at
finite {\bf k} mixes different spin states. This mechanism, known as the
Elliot-Yafet mechanism \cite{EY}, was found to be the dominant spin flip
mechanism in exciton spin flip in GaAs quantum wells
\cite{snoke-ruehle}. In quantum dots, however, this effect cannot occur at low temperature, because
the carriers are confined to the lowest quantized state in the well. Scattering
between different states can occur only along with jumps in energy which are
large compared to $k_BT$. One therefore expects spin flip to be greatly
suppressed in quantum dots.

Measurements reported here, however, as well as recent measurements by other means \cite{yugo1},
imply a much shorter spin flip time for excitons in InP quantum dots, much less than a nanosecond.
This result is not related to coupling between the dots, since they are known to be well isolated,
but may be related to the geometry  of the dots.

{\bf Experimental Method}. As is well known, not all states in semiconductors couple directly to
the optical field. In bulk semiconductors and quantum wells, there are many examples of ``dark''
states for which light emission is forbidden in first order due to symmetry, in contrast to
``bright'' states which have a dipole-allowed optical matrix element. To convert from a dark state
to a bright state involves a spin flip, because the angular momentum of the two states is
different.

In previous work \cite{snoke-ruehle}, the
spin flip time from dark to bright excitons in GaAs quantum wells was measured
by exciting the dark states by two-photon absorption and detecting the
single-photon luminescence from bright excitons. Because the bright excitons
were not excited directly by two-photon excitation, the rise time of the
luminescence from the bright excitons gave a direct measurement of the spin
flip time for the conversion process.

In this work we apply the same procedure to quantum dots of another
III-V material, InP. A priori, it is not obvious that dark states exist in the dots, because the
change in symmetry properties may mean that all the confined states have dipole coupling to the
light field. If they do exist, we should observe a rise time of the luminescence following
the laser pulse which creates the dark excitons, just as in the case of dark states in quantum
wells.  Having proven the existence of dark states in this way, we can then determine the spin flip
time for conversion from dark to bright states.

The quantum dots used for these experiments were a single layer of 3.0 ML InP quantum dots, with a
nominal height of 3.8 nm and diameter 15.7 nm and dot density $3.2\times 10^{10}$ cm$^{-2}$. The
gap energy of the InP is nominally 1.42 eV at low temperature. The dots are enclosed by
Ga$_{0.52}$In$_{0.48}$P barriers, which have band gap of 1.91 eV.  As shown in earlier experiments
\cite{zundel}, the luminescence from the confined states of the dots occurs at photon energy of
1.805 eV, with a full width at half maximum of 0.041 eV due to the distribution of the dot size.

The symmetry properties of the quantum dots are related to
those of quantum wells of the same material. Quantum wells of III-V semiconductors belong to the
$D_{2d}$ symmetry group. The topmost valence band, or heavy hole band, has $\Gamma_7$ symmetry in
this group. The lowest energy excitons, created from $\Gamma_6$ conduction electrons and heavy
holes, are split into an optically-active $\Gamma_5$ doublet and two optically inactive ``dark''
exciton states with $\Gamma_3$ and $\Gamma_4$ symmetry, i.e. basis states $| j = 2, m = 2 \rangle
\pm i| j = 2, m = -2\rangle$. Previous work \cite{magres} has shown that these dark states
are split from the bright states by energies on the order of 100
$\mu$eV in III-V semiconductor quantum wells. 

The symmetry of the quantum well is further lowered in the quantum dots, so that all of these
states will become nondegenerate. Nevertheless, the quantum states
in the dot will have character similar to the states of the quantum well of the same material.
Single-dot spectroscopy in magnetic field has shown that the states are split into two pairs at
with bright and dark character, with state splitting of a few hundred
$\mu$eV for dots made of a wide range of III-V and II-VI materials, including GaAs
\cite{ivchenko,phillips}, InAs \cite{bayer}, CdSe
\cite{nirmal,efros}, and CdTe \cite{bes}. The splitting of the degenerate bright 
states can also be of the order of a hundred $\mu$eV \cite{sugi}. The splitting of the dark and
bright states in InP dots similar to ours has recently been estimated at less than 30 $\mu$eV
\cite{yugo2}.

The dots were excited by means of two-photon
absorption using a Ti:Sapphire-pumped OPO with photon energy of 0.9 eV, or 1375
nm. We used a microscope objective to focus the laser to a spot size of
approximately 6 $\mu$m, and a laser pulse energy of around 1.3  nJ (100 mW at
80 MHz repetition rate) in order to obtain a good signal too noise ratio. The
luminescence was recorded using a Hamamatsu streak camera with a S-1 cathode and a temporal
resolution of about 10 ps.

The greatest challenge in studying the transient optical signal from
the dots on the GaAs substrate is ensuring that the signal arises from direct laser excitation of
the dots and not from carriers excited in the substrate which find their way to the dots. This
latter process can certainly occur, as we have established by the observation that luminescence
from the dots occurs even when the two-photon excitation energy is well below the lowest excited
state of the dots. Substrate carriers can enter the dots by a two-step process.
First, two-photon absorption can occur in the substrate, leading to
free carriers, and these carriers can be excited by absorption of a third photon to energies well
above the barrier height of the quantum dots. Evidence for this comes from the strong luminescence
signal from the GaAs substrate which includes a tail to very high energy at early times, during the
laser pulse. Some fraction of these hot carriers can then excite the quantum dots. This can occur
either if hot carriers diffuse across the barriers, or if luminescence
photons from the hot carriers are reabsorbed by the dots.

If this process is the dominant source of the signal, then we can not say anything about dark states
in the quantum dots, because the spin of the carriers will presumably be randomized during the
migration process into the dots. The rise time of the luminescence signal will give us information
only about the relaxation processes which lead to this indirect excitation process.

One way to distinguish between an effect like this and true two-photon
excitation of the quantum dots uses the fact that the two-photon excitation
process has a resonance at the energy of the quantum confined states, while the
hot-carrier process of exciting the dots is relatively insensitive to the
wavelength of the exciting light. Fig. 1 shows the total intensity of the
luminescence from the dots on a logarithmic scale as a function of the laser wavelength. The
increase towards 1345 nm corresponds to photons with energy one half the energy of the
quantum dot luminescence.  As seen in this figure, well below the dot
resonance, the luminescence signal from the hot carrier effect is approximately
constant. We can assume that the excess signal near the resonance arises from
direct two-photon excitation.

Another way to distinguish between the different processes is to note that they
will have different power laws. The intensity of the signal from two-photon
excitation should be proportional to the square of the laser power, while the
intensity of the signal from indirect pumping of the dots hot carriers in the
substrate has a much stronger intensity dependence. Fig. 2 shows a comparison
of the total luminescence intensity from the dots in the case of excitation at
two different wavelengths. In the case of excitation at the dot resonance, the
intensity dependence fits a power law of $I^{2.45}$, approximately equal to the
expected power law of $I^2$. In the case of excitation well below the
resonance, the power law fits a dependence of $I^{5.5}$, which is much
stronger. A power law of $I^3$ would be expected for a straightforward
three-photon process in which carriers created in the substrate absorption by
two-photon absorption were excited into the dots by absorption of a third
photon. The stronger power dependence reflects the fact that at high densities,
the ``hot phonon'' effect strongly reduces carrier cooling in the substrate. This
hot phonon effect was studied in detail several years ago: carrier cooling 
at high carrier densities is reduced since the optical phonons preferentially
emitted at high carrier energy have finite lifetime. A nonthermal optical
phonon occupation is built up very fast and these hot phonons strongly
reabsorbed at high densities leading to a reduction of the net energy flow from
the carrier into the lattice system \cite{zhouPRB92}.  This is verified by the fact that the
high energy tail of the substrate luminescence becomes much stronger at high
excitation density, and at the highest excitation density a substantial
fraction of the substrate luminescence additionally overlaps the luminescence
spectrum of the dots immediately after the laser pulse. The fact that the
exponent in the case of resonant excitation is 2.45 instead of exactly 2 is
likely due to the fact that the signal in this case is a sum of both the direct
two-photon excitation signal and the signal from carriers indirectly excited
from the substrate.

The difference in the power laws allows us to pick an excitation
regime in which the signal from
direct two-photon excitation is much stronger than that from indirect
transfer of carriers from the
substrate. As seen in Fig. 2, at 60 mW average power, the signal from
the direct two-photon excitation
process is more than a factor of ten greater than the signal from
excitation of the substrate. We therefore
excite the sample with laser power in this regime, instead of the
highest possible laser power,
in order to maximize the signal from direct two-photon excitation.

{\bf Results}. Our observations indicate that there is a clear rise time of
the luminescence following the nearly resonant excitation by the
laser pulse. The risetime is consistent with the expected behavior for conversion of dark to bright
excitons, which confirms the existence of dark states in the quantum dots. Fig. 3 shows the total
luminescence intensity from the dots as a function of time, at $T= 10$ K, for laser
power 60 mW which, as discussed in the
previous section, is low enough that the effects from hot carriers in
the GaAs substrate should be negligible.

The fit to the data of Fig. 3 is a solution of a simple set of rate equations,
\begin{eqnarray}
\frac{\partial n_1}{\partial t} &=& -\frac{n_1}{\tau_{12}} +
\frac{n_2}{\tau_{21}} - \frac{n_1}{\tau_0}
\nonumber\\
\frac{\partial n_2}{\partial t} &=& \frac{n_1}{\tau_{12}} -
\frac{n_2}{\tau_{21}}, \label{rate}
\end{eqnarray}
where $n_1$ and $n_2$ are the number of bright and dark excitons, respectively.
The decay time $\tau_0$ is the radiative lifetime of the excitons in the bright
state, while $\tau_{12}$ and $\tau_{21}$ are times for conversion from bright
states to dark and from dark to bright, respectively. Since both the bright and dark states are
doublets, we assume equal degeneracy for both states.
In principle, $\tau_{12}$ and $\tau_{21}$ can be different, as found for
quantum wells at low temperature \cite{snoke-ruehle}, but if the energy
splitting between the states is small compared to $k_BT$, then these rates will
be nearly the same. In the present experiments, the temperature ranged from 10
K to 75 K. Assuming that the splitting of the dark and bright states is of the
order of 100 $\mu$eV or less, as discussed above, the splitting is
much less than $k_BT$, and therefore we can set $\tau_{12} = \tau_{21} = \tau$. In this case, the
solution of the equations (\ref{rate}) for the initial condition $n_1(0) = 0, 
n_2(0) = 1$, is
\begin{eqnarray}
n_1(t) &=& \frac{\tau_0}{\sqrt{\tau^2 +
4\tau_0^2}}\left(e^{-t\left(\tau+2\tau_0 -\sqrt{\tau^2 +
4\tau_0^2}\right)/2\tau\tau_0} - e^{-t\left(\tau+2\tau_0 + \sqrt{\tau^2 +
4\tau_0^2}\right)/2\tau\tau_0}   \right) \label{dr}\\
&\equiv& C\left(e^{-t/\tau_d} - e^{-t/\tau_r} \right)\nonumber
\end{eqnarray}

Surprisingly, this solution implies that the minimum ratio of the
decay time
$\tau_d$ to the rise time $\tau_r$ is $5.85 = (2 + \sqrt{2})/(2 - \sqrt{2})$, for the case
$\tau = 2\tau_0$. For all other choices of the time constants, the ratio of
the decay time to the rise time is larger than this. This shows that it is improper to
interpret the rise time of the luminescence as the spin flip time and the decay
time as the radiative lifetime. Because of the interconversion between the
states, both time scales depend on both $\tau$ and $\tau_0$.
Within the experimental uncertainty, our data at low $T$ give this ratio, which
implies that $\tau \simeq 2\tau_0$, and in general, that $\tau$ is longer than
$\tau_0$.

When the ratio deviates from this minimum value, there are two possible solutions for $\tau$ and
$\tau_0$ given the experimental values of $\tau_r$ and $\tau_d$, as follows:
\begin{eqnarray}
\tau_0 = \frac{\tau_d +\tau_r - \sqrt{\tau_d^2 - 6\tau_d\tau_r +\tau_r^2}}{4} &,&   \ \
\tau =  \frac{\tau_d +\tau_r + \sqrt{\tau_d^2 - 6\tau_d\tau_r +\tau_r^2}}{2} \label{solution}\\
&\mbox{or}\nonumber \\
\tau_0 = \frac{\tau_d +\tau_r + \sqrt{\tau_d^2 - 6\tau_d\tau_r +\tau_r^2}}{4} &,& \ \
\tau =  \frac{\tau_d +\tau_r - \sqrt{\tau_d^2 - 6\tau_d\tau_r +\tau_r^2}}{2} .
\label{sol}
\end{eqnarray}
We take the former solution here, which is consistent with the radiative recombination time $\tau_0$
essentially independent of the temperature, as expected, and the spin flip time $\tau$ longer than
$\tau_0$ in all cases. Table 1 gives the temperature dependence of the values deduced from these
fits. As seen in this table, the conversion time from dark to bright states remains approximately
200 ps at low temperature. If the alternate solution (\ref{sol}) is taken, then the implied spin
flip time drops to tens of picoseconds while the radiative recombination time
becomes significantly longer at $T=75$ K.

We want to stress that the model presented here implies that even in the case of resonant
single-photon excitation, the decay of the dot luminescence is dominated by the interconversion of
the dark and bright states. Long after the laser pulse, the luminescence decay time for
single-photon excitation will be the same as that given in Equation (\ref{dr}) for two-photon
excitation. According to this equation, the decay time $\tau_d$ of the photoluminescence will be
equal to
$2\tau_0$ in the limit that the spin flip time is much shorter than the radiative lifetime; in the
limit of long spin flip time, the decay time
$\tau_d$ will be essentially equal to the spin flip time $\tau$, not $\tau_0$.  This effect of
interconversion of bright and dark excitons, although discussed in detail for the case of quantum
wells
\cite{vinattieri}, has been neglected in several previous publications; for example, in Refs.
\cite{ruehleNC95,paillardAPL00,bayerPRL01} the photoluminescence decay time was taken simply as the
radiative decay time while as seen here, the decay rate even at late times is in general a function
of the spin flip time from dark states.

The radiative decay rates found here are consistent with the strong localization of the carriers
which entends the recombination time. Typical intrinsic decay times for quantum wells are a few tens
of picoseconds \cite{andreani,deveaud}, while the recombination time found for the dots in these
experiments is around 70 ps, which corresponds to a total radiative decay time of 140 ps in the
limit of fast spin flip.

We find no power dependence of the spin flip time $\tau$. This is not surprising, since the
excitation density is so low that it is unlikely that there is more than one
electron per dot. In this case, each dot relaxes individually.

The most likely reason for the variation in the time constants at $T=75$K is that at high
temperature, carriers are excited into higher quantized states, which lie approximately 10 meV
above the lowest state \cite{ulrich} so that our simple two-state model breaks down. At low
temperature, the time constants are essentially independent of temperature. 

The short spin flip time is surprising, because as
discussed above, previous studies have found a dramatically slower rate for spin flip
in quantum dots at low temperature. As mentioned above, however, another study of InP dots
\cite{yugo1} has found a very short time constant for depolarization of the luminescence
from the dots following excitation with circular polarized light. This short lifetime was
interpreted by the authors of \cite{yugo1} as due to interference of the light emitted from the
ensemble of quantum dots with large inhomogeneous broadening. That explanation does not apply to
the experiments reported here, however, because our method of measuring the time scale for spin flip
from dark to bright excitons is insensitive to the inhomogeneous broadening of the ensemble. A
possible explanation for the large range of spin flip times may come from the differences in
geometry of the dots. Woods, Reinecke, and Lyanda-Geller \cite{woods} have calculated the rate of
spin flip in dots as a function of the dot geometry, for two possible mechanisms, acoustic phonon
emission and interface ripples, analogous to surface acoustic waves on the interface between the
dots and the barriers. They found a very strong size dependence; in particular, the height of our
quantum dots of 3.8 nm lies in the range at which they found a strongly increasing rate of spin
flip with decreasing size. 

Dark states play an important role in the relaxation of bright excitons in quantum
dots, controlling the observed luminescence decay rate. We find a nearly constant rate of
conversion from bright to dark excitons in InP quantum dots at low temperature, approximately 200
ps. The results from InP show that it cannot be generally assumed that spin flip times are always
long in quantum dots.

\newpage

\begin{table}
\begin{tabular}{c|cccc}
& 10 K & 20 K & 40 K & 75 K \\ \hline
rise time (ps) & $55 \pm 9$ & $53 \pm 10$ & $54 \pm 8$ & $45 \pm 8$\\
decay time (ps)  & $297 \pm 15$ & $299 \pm 16$ & $349 \pm 17$ & $502 \pm 22$\\
$\tau$ (ps) & $209 \pm 27$  & $218 \pm 31$ & $255 \pm 44$ & $445 \pm 36$ \\
$\tau_0$ (ps) & $71 \pm 9$ & $68 \pm 13$ & $73 \pm 16$ & $50 \pm 11$
\end{tabular}
\caption{Time constants determined by the fits of the data to the model discussed in the text. The
ranges of the values of $\tau$ and $\tau_0$ given here represent all possible
solutions of (\ref{solution}) using values of $\tau_r$ and $\tau_d$ which fall within the ranges of
uncertainty of their fit values.}
\end{table}

\begin{figure}
\vspace{.5cm}
\hspace{2cm}
\epsfxsize=.9\hsize \epsfbox{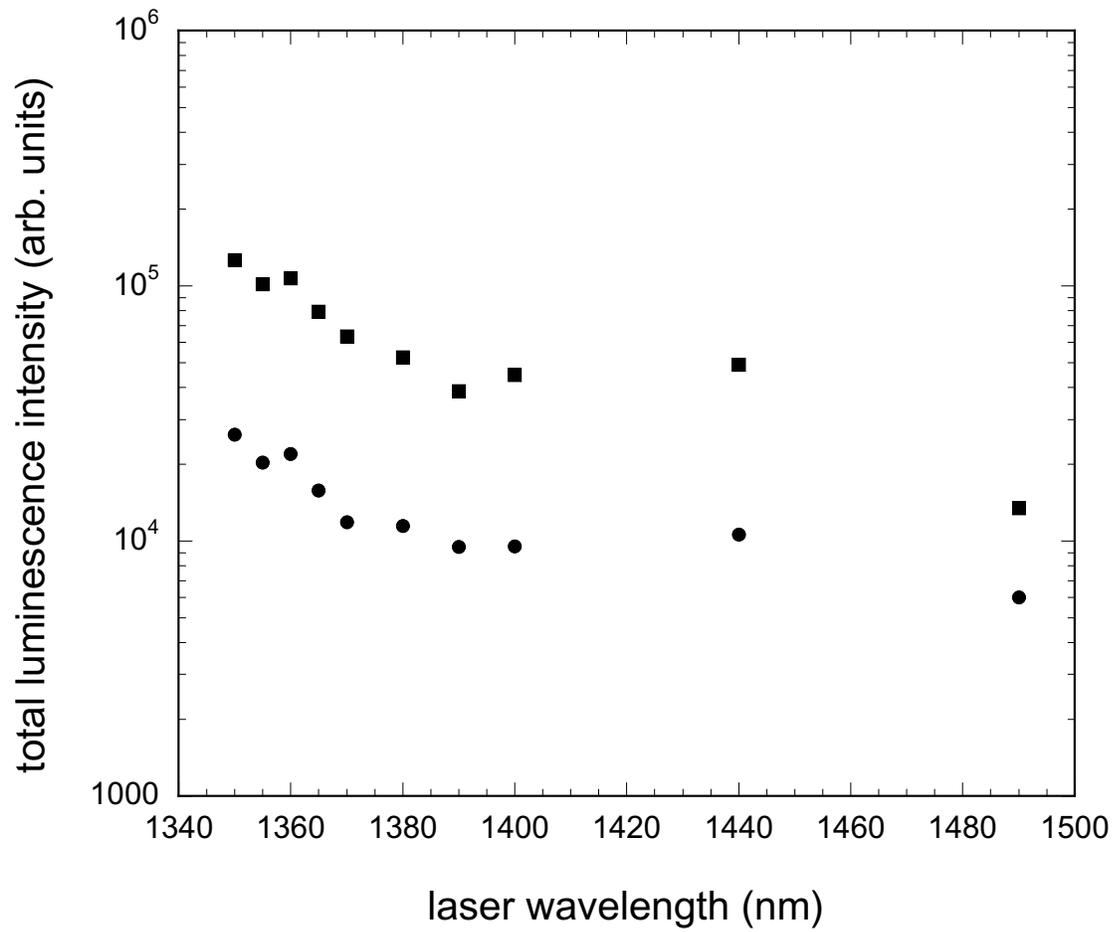}
\vspace{.5cm} \caption{Total luminescence
intensity from the dots as a function of the excitation laser wavelength.}
\end{figure}

\begin{figure}
\vspace{.5cm} \hspace{2cm}
\epsfxsize=.9\hsize \epsfbox{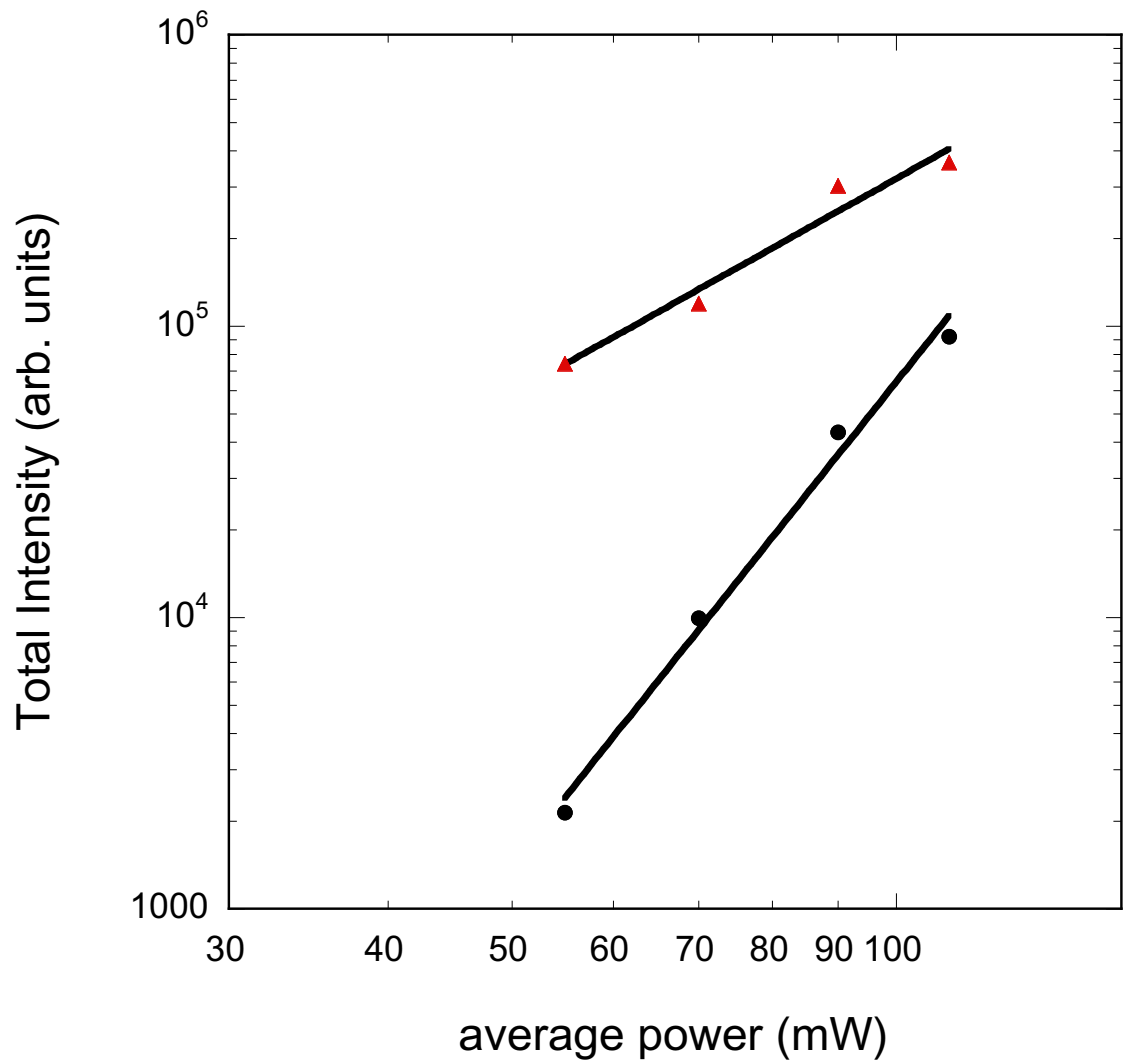}
\vspace{.5cm} \caption{Total luminescence
intensity as a function of average laser power, for two different
wavelengths. Squares: 1465 nm. Triangles: 1340 nm.  }
\end{figure}

\begin{figure}
\vspace{.5cm}
\hspace{2cm}
\epsfxsize=.9\hsize 
\epsfbox{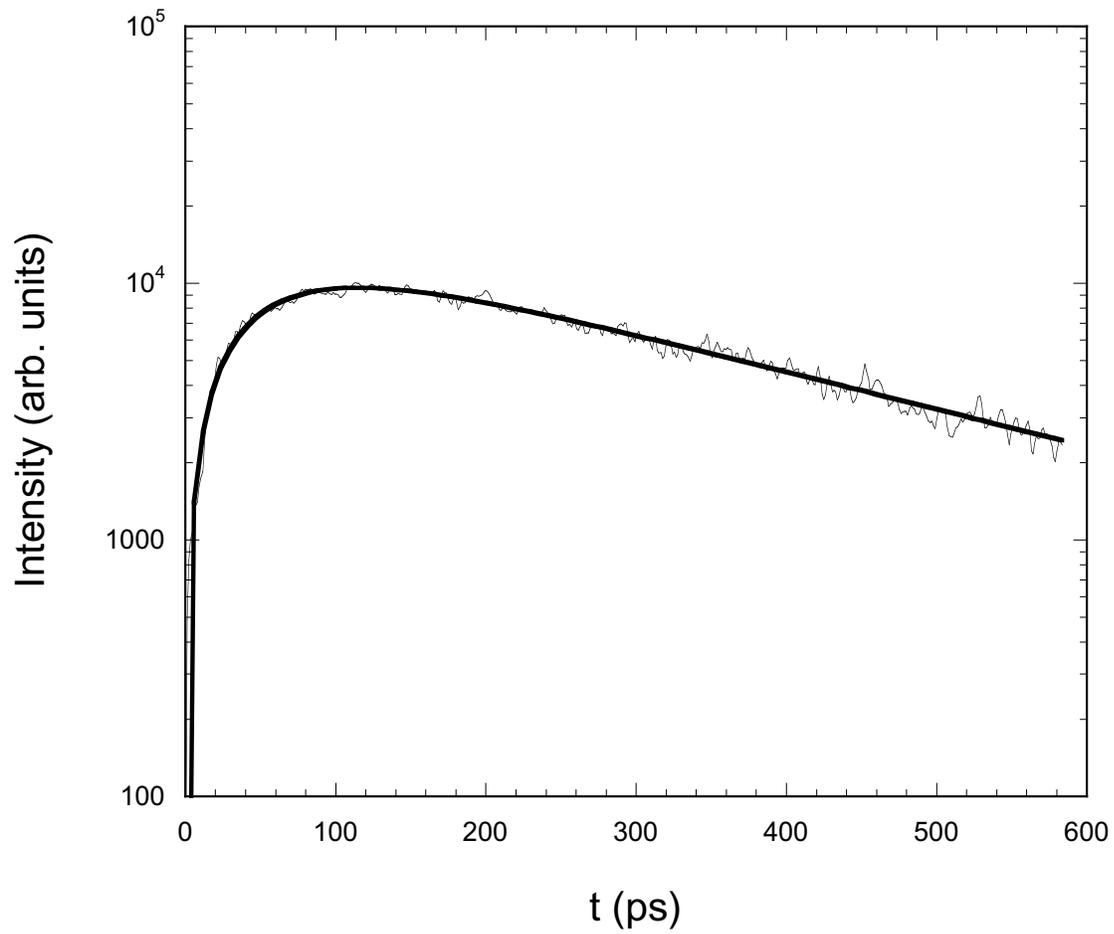}
\vspace{.5cm} 
\caption{Total
luminescence intensity from the dots as a function of time, following
two-photon excitation by a laser pulse at $t=0$. Heavy line: fit to the
theory discussed in the text.}
\end{figure}

\end{document}